\documentclass[useAMS,usenatbib]{mn2e}
\usepackage{times,aas_macros}
\usepackage{graphicx}
\usepackage{amssymb}
\usepackage{amsmath}
\usepackage{rotating}
\input{epsf}

\title[X-ray Lags in PDS~456]
{X-ray Lags in PDS~456 Revealed by \emph{Suzaku} Observations}

\author[C.-Y. Chiang et al.]
{Chia-Ying Chiang$^{1}\thanks{E-mail: ft8320@wayne.edu}$,
E. M. Cackett$^{1}$,
A. Zoghbi$^{2}$,
A. C. Fabian$^{3}$, 
E. Kara$^{4,5,6}$,
\newauthor M. L. Parker$^{3}$, 
C. S. Reynolds$^{4,6}$, 
and D. J. Walton$^{3}$
\\
$^{1}$Department of Physics \& Astronomy, Wayne State University, Detroit, MI 48202, USA\\
$^{2}$Department of Astronomy, The University of Michigan, 500 Church Street, Ann Arbor, MI 48109-1046, USA\\
$^{3}$Institute of Astronomy, University of Cambridge, Madingley Road, CB3 0HA\\
$^{4}$Department of Astronomy, University of Maryland, College Park, Maryland 20742, USA\\
$^{5}$X-ray Astrophysics Laboratory, NASA/Goddard Space Flight Center, Greenbelt, Maryland 20771, USA\\
$^{6}$Joint Space Science Institute, University of Maryland, College Park, Maryland 20742, USA\\
}

\pubyear{2017}

\begin{document}
\label{firstpage} 
\topmargin = -1.0cm
\pagerange{\pageref{firstpage}--\pageref{lastpage}} 
\maketitle

\begin{abstract}
X-ray reverberation lags from the vicinity of supermassive black holes have been
detected in almost 30 AGN. The soft lag, which is the time delay between the hard
and soft X-ray light curves, is usually interpreted as the time difference between the
direct and reflected emission, but is alternatively suggested to arise from the direct 
and scattering emission from distant clouds. By analysing the archival \emph{Suzaku} 
observations totalling an exposure time of $\sim$ 770 ks, we discover a soft lag of 
$10\pm3.4$~ks at $9.58\times10^{-6}$ Hz in the luminous quasar PDS~456, which is 
the longest soft lag and lowest Fourier frequency reported to date. In this study, 
we use the maximum likelihood method to deal with non-continuous nature of the  
\emph{Suzaku} light curves. The result follows the mass-scaling relation for soft lags, 
which further supports that soft lags originate from the innermost areas of AGN and 
hence are best interpreted by the reflection scenario. Spectral analysis has been 
performed in this work and we find no evidence of clumpy partial-covering absorbers. 
The spectrum can be explained by a self-consistent relativistic reflection model with 
warm absorbers, and spectral variations over epochs can be accounted for by
the change of the continuum, and of column density and ionization states of the 
warm absorbers.

\end{abstract}


\section{Introduction}

The geometry of an Active Galactic Nuclei (AGN) is widely accepted to be 
an accretion disc rotating around the central supermassive black hole, and 
illuminated by a hard X-ray source known as the `corona'. The accretion disc 
absorbs high-energy photons, and then atomic transitions take place and emit 
a series of X-ray emission lines in which the Fe K$\alpha$ line is the most 
prominent. The fluorescent lines are intrinsically narrow but sometimes appear 
broad and asymmetric in the observed spectrum. The shape of the broad, skewed 
Fe K$\alpha$ emission line discovered in the Type 1 AGN MCG-6-30-15 
\citep{Tanaka95} can be caused by special and general relativistic effects \citep{Fabian89} 
if the line originates from the innermost part of the accretion disc where strong
gravitational fields of the central black hole cast important effects. This provides a 
powerful method of measuring the inner radius of the accretion disc (and hence 
the black hole spin in accreting black hole systems) using the iron line profile 
\citep[see reviews by][]{Miller07,Fabian10}. Nevertheless, some researchers 
suggested that the central engine cannot be directly observed even in the X-ray band
and proposed a partial-covering scenario to explain the spectral properties of the X-ray
spectrum of AGN \citep{Inoue03,Miller09}. If this is the case, we cannot obtain
information from the innermost areas of black holes via X-ray observations. Thus, the inner radius
of the accretion disc cannot be measured and alternative methods need to be developed
to probe the vicinity of accreting systems.
\citet{Risaliti13} used the simultaneous
\emph{XMM-Newton} and \emph{NuSTAR} observations to break spectral degeneracy and found that reflection
of NGC~1365 arises from a region within 2.5 gravitational radii ($R_{\rm g}=GM/c^2$), 
but it is difficult to distinguish the relativistic reflection model and the partial-covering 
scenario using spectral analysis in most X-ray observations \citep{Walton14}. In order to break 
this degeneracy, detailed spectral-timing analyses have been performed in recent years.  
The principle goal of these studies have been to search for the reverberation lags predicted 
by the disk reflection scenario \citep{Fabian89,Reynolds99}.

The first significant X-ray reverberation lag was found in the narrow-line Seyfert 1
(NLS1) galaxy 1H0707-495 \citep[][see \citet{McHardy07} for an earlier hint in Ark~564]
{Fabian09}. A $\sim$30 s soft lag, which 
implies that the soft X-rays lag behind the hard X-rays by 30 seconds, was
reported as evidence supporting the reflection scenario. Similar soft lags
have been reported in a number of AGN as well \citep{Emm11,Zoghbi12,Zoghbi14,
Cackett13,DeMarco13,Kara13,Kara13Ark,Kara15,Kara16}. 
However, alternative explanations of soft lags were proposed. \citet{Jin13}
suggested that the majority of the soft excess in the NLS1 PG~1244+026 is produced by
a cool Comptonization component rather than reflection. \citet{Gardner14} further examined
the source and found that a model for the soft excess consisting of intrinsic emission from
the accretion flow and reprocessed emission can explain the spectral and timing
properties of PG~1244+026.
Some work suggested that soft lags 
are caused by scattering reverberation from distant (tens to hundreds of 
$R_{\rm g}$) material away from the primary 
X-ray source \citep{Miller10,Legg12,Turner17}. These papers suggested that
the true signature of reverberation is the low-frequency lags, and the high-frequency lags 
inferred as reverberation is just phase-wrapping of the low-frequency signal \citep[however,
see counter-arguments in][]{Zoghbi11}. 
\citet{Kara13Ark} then showed that the lag-energy spectra of these frequency regimes were 
distinct, demonstrating the need for different physical processes.
In addition, \citet{Walton13} found a low-frequency lag in NGC~6814 
where the spectrum seems to be dominated by only the powerlaw continuum. 
Both works argued against the distant reverberation scenario, supporting an inner disk reflection
reverberation origin for the high-frequency lags. \citet{DeMarco13} 
examined a sample of AGN with black hole masses spanning 
$\sim10^{6}-10^{8}M_{\odot}$ and found that more massive AGN have
larger soft lags appearing at lower Fourier frequencies. The mass-scaling
relation and values of lags indicate that soft lags originate in the innermost 
area of AGN.
The discovery of Fe K$\alpha$ reverberation in NGC 4151 \citet{Zoghbi12}
and 20 subsequent AGN 
\citep{Zoghbi13Fe,Fabian13,Kara13IRAS,Kara13,Kara13Ark,Kara14,Kara15,Kara16}
further supports reverberation from the inner disc (see \citet{Uttley14} for a detailed 
review of X-ray reverberation). The self-consistency of the relativistic reflection
scenario indicates that X-ray observations provide information from the vicinity
of black holes and can help understand the central engines of AGN.

PDS~456 is a luminous ($L_{\rm bol}=10^{47}$ erg s$^{-1}$, \citealt{Simpson99}) 
nearby (z = 0.184, \citealt{Torres97}), radio-quiet quasar, which is known for powerful disc 
winds/outflows \citep{Reeves03,Reeves09} and rapid X-ray variability
\citep{Reeves00,Reeves02}. The source was observed several times with major X-ray observatories
including \emph{BeppoSAX} \citep{Vignali00}, \emph{ASCA} \citep{Vignali00}, 
\emph{XMM-Newton} \citep{Nardini15}, \emph{Suzaku} \citep{Reeves14}, 
and \emph{NuSTAR}.
\citet{Vignali00} and \citet{Reeves00} both reported a significant absorption 
feature at around $8-9$ keV rest frame 
in an early \emph{ASCA} observation. The 2007 \emph{Suzaku} observation further
revealed that the feature is likely the blue-shifted 6.97 keV iron absorption line 
at velocities of $\sim0.25c$ \citep{Reeves09}. The ultrafast outflow (UFO) has been 
confirmed in later observations as well \citep{Reeves14,Gofford14,Nardini15,
Matzeu16}. 
UFOs have garnered wide interest because these energetic outflows are capable of 
contributing to AGN feedback. \citet{Nardini15} interpreted the broadened Fe K$\alpha$ 
emission line with the adjacent absorption feature caused by the UFO in
\emph{XMM-Newton}/\emph{NuSATR} observation of PDS~456 as a 
P-Cygni profile produced by an expanding gaseous envelope. In their analysis 
they also tested the disc absorption scenario \citep{Gallo11,Gallo13}, but found that 
it is not required statistically.
Apparent spectral variability can be seen by 
comparing observations taken at different epochs and different observatories. Some previous
work used a partial-covering model to account for most of the spectral variability 
\citep{Gofford14,Nardini15,Matzeu16}. We revisit the long, archival \emph{Suzaku}
observations to examine if partial-covering absorbers are required to explain
the spectrum. Based on the archival observations, the 
source shows short-term variability and hence an ideal source for timing analysis. PDS~456
has a high-mass black hole (Log $M_{\rm BH}=8.91\pm0.50$, \citealt{Zhou05}), beyond
the mass range of the mass-scaling relation in \citet{DeMarco13}.
In this work, we use the \emph{Suzaku} observations to examine 
whether a soft lag(s) can be found in this source, such as other AGN with lower 
masses, and if the result still follows the mass-scaling relation. 
With the spectral-timing analysis, we re-examine the data to figure out if
the relativistic reflection model gives consistent results.

\begin{table}
  \centering
  \caption{Summary of the \emph{Suzaku} observations of PDS~456. }
  \label{summary}
  \begin{tabular}{clcc}
  \hline
  \hline
Obs. ID &  Start Date & Exposure (ks) & Count Rate (ct s$^{-1}$)\\
\hline
701056010 & 2007 Feb. 24 & 190.6   &  $(2.7\pm0.1)\times10^{-1}$ \\
705041010 & 2011 March 16& 125.5   & $(1.4\pm0.1)\times10^{-1}$  \\
707035010 & 2013 Feb. 21 & 182.2  & $(6.7\pm0.1)\times10^{-2}$  \\
707035020 & 2013 March 03& 164.8   & $(4.4\pm0.1)\times10^{-2}$   \\
707035030 & 2013  March 08& 108.2  & $(5.2\pm0.1)\times10^{-2}$    \\
\hline
\hline
\end{tabular}
\end{table}

\begin{figure}
\begin{center}
\leavevmode \epsfxsize=8.5cm \epsfbox{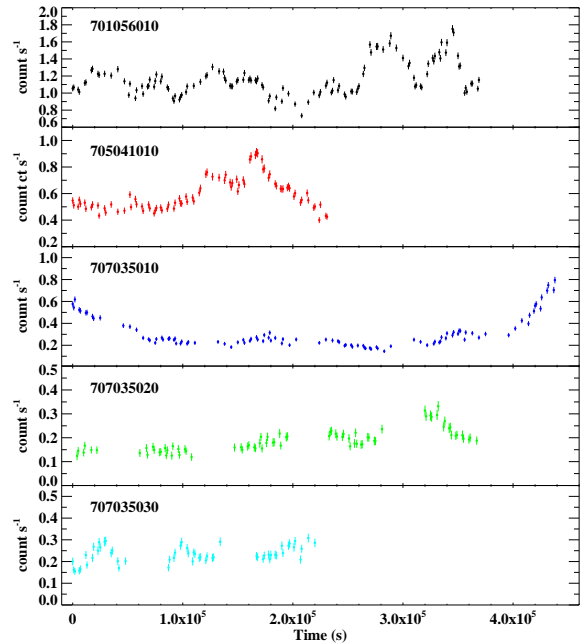}
\end{center}
\caption{The 0.3-10.0 keV XIS (XIS0, XIS1, and XIS3 combined) light 
curves for all observations. It can be seen that the duration of the longest observation is $\sim$
438 ks, which is much longer than an \emph{XMM-Newton} orbit ($\sim$ 130 ks). 
All the light curves are highly variable.
}
\label{lc}
\end{figure}

\section{Data Reduction} \label{section_da}

PDS~456 was observed with \emph{Suzaku} in 2007 (Obs ID: 701056010), 
2011 (Obs ID: 705041010), and 2013 (Obs IDs: 707035010, 707035020, and
707035030), with good exposure times of $\sim$ 190.6 ks, $\sim$ 125.5 ks 
and $\sim$ 455.2 ks, respectively (see Table \ref{summary}). The X-ray Imaging Spectrometer (XIS) was 
operated in the normal mode in all observations. The data were reduced using the 
{\sevensize HEASOFT} V6.16 following the Suzaku Data Reduction Guide. We 
used a circular region with a 150\arcsec radius to extract source spectra and light
curves; the same shaped region was used to extract background products in a source-free region. 
Response files were generated by the {\sevensize XISRESP} script. We combined
spectra and response files of the front-illuminated (FI) CCD XIS detectors (XIS0 and 
XIS3) using {\sevensize ADDASCASPEC} in {\sevensize FTOOL}. Spectra of different
Obs IDs of the 2013 observation were combined using {\sevensize ADDASCASPEC}
as well. The total absorption-corrected $0.5-10.0$ keV FI-detectors fluxes are 
$9.2\times10^{-12}$ erg cm$^{-2}$ s$^{-1}$, $6.9\times10^{-12}$ erg cm$^{-2}$ 
s$^{-1}$, and $4.1\times10^{-12}$ erg cm$^{-2}$ s$^{-1}$ for the 2007, 2011 and 2013 
observations. We produced $0.3-10.0$ keV background-corrected light curves with 
1 ks time bin for the XIS0, XIS1, and XIS3 detectors, and combined all of them to form 
an overall light curve for each observation. The light curves are displayed in Fig. \ref{lc}
and it can been seen that the source shows variability in all observations.

The Hard X-ray Detector (HXD) was operated in XIS nominal pointing mode in all 
observations. We extracted the non-X-ray background using the tuned model. The 
total background was generated using the PIN reduction script 
{\sevensize HXDPINXBPI}, which combines the non-X-ray background and cosmic 
background automatically. The net count rates are 0.007 count s$^{-1}$, 0.001 count 
s$^{-1}$, and 0.003 count s$^{-1}$ in the effective PIN energy band $15.0-45.0$ keV,
which are 2.0\%, 0.8\% and 1.1\% of the total background count rate. The source 
was fairly faint in the band during the later two observations, and only marginally 
detected in the 2007 observation. We did not include the PIN light curves in the 
following timing analysis, and only used the 2007 data for spectral analysis.

\begin{table*}
 \caption{Parameters for the best-fitting reflection model to the data,
 in which $N_{\mbox{\scriptsize H}}$ is given in $10^{21}$ cm$^{-2}$, and $\xi$ in 
 erg cm s$^{-1}$. The normalisation of the powerlaw component is expressed in 
 photons keV$^{-1}$ cm$^{-2}$ s$^{-1}$.  The hard upper limit of $q$ is 8, and that of 
 $A_{\rm Fe}$ is 10.}
\label{spec}
\centering
\begin{tabular}{@{}llccc}
\hline\hline
Component & Parameter & 2007 & 2011 & 2013 \\
\hline
TBABS & $N_{\rm H}$ ($10^{21}$ cm$^{-2}$)  & $2.6\pm0.2$ & $-$ & $-$\\
XSTAR$_{1}$ (UFO) & $N_{\rm H}$ ($10^{22}$ cm$^{-2}$) & $4.1^{+1.2}_{-3.0}$ & $0.9^{+0.9}_{-0.5}$  & $1.8^{+0.8}_{-0.6}$\\
& Ionization parameter, log $\xi$ & $4.3^{+0.2}_{-0.4}$ & $3.5^{+1.0}_{-0.1}$ & $3.5\pm0.1$ \\
& Redshift, $z$ & $-(9.2^{+0.6}_{-0.5})\times10^{-2}$ & $-$ & $-$\\
XSTAR$_{2}$ (low-energy) & $N_{\rm H}$ ($10^{21}$ cm$^{-2}$) & $-$ & $3.4^{+1.5}_{-0.9}$  & $4.8^{+2.1}_{-1.5}$\\
& Ionization parameter, log $\xi$ & $-$ & $0.5^{+0.2}$ & $1.2\pm0.1$ \\
& Redshift, $z$ & $-$ & $(0.184)$ & $(0.184)$\\
POWERLAW & Photon index, $\Gamma$ & $2.35\pm0.01$ & $2.17^{+0.08}_{-0.06}$ & $1.73^{+0.06}_{-0.05}$ \\
 & Norm & $(2.2\pm0.1)\times10^{-3}$ & $(1.2\pm0.1)\times10^{-3}$ & $(3.5^{+0.3}_{-0.2})\times10^{-4}$\\
RELCONV & emissivity index, $q$ & $2.2^{+0.2}_{-0.1}$ & $5.3^{+0.7}_{-0.9}$ & $6.1^{+0.4}_{-0.5}$ \\
 & Spin parameter, $a^{*}$ & $>0.990$ & $-$ & $-$\\
 & Inclination, $i$ (deg) & $(65\pm2)^{\circ}$ & $-$ & $-$ \\
REFLIONX & Iron abundance / solar, $A_{\rm Fe}$ & $> 9.3$ & $-$ & $-$\\
 & Ionization parameter, $\xi$ & $67^{+19}_{-7}$  & $70^{+30}_{-10}$ & $120^{+30}_{-10}$ \\
 & Norm & $(1.0\pm0.3)\times10^{-6}$ & $(1.9^{+0.9}_{-0.6})\times10^{-6}$ & $(9.4^{+1.6}_{-1.4})\times10^{-7}$ \\
$\chi^{2}/d.o.f.$ & & 1131/1107 & 784/786 & 1168/1122\\
\hline\hline\\
\end{tabular}
\end{table*}

\section{Data Analysis}

We perform both spectral and timing analyses, described below. \citet{Reeves09}
claimed that a partial-covering Compton-thick absorber is needed to explain the hard
excess above 10 keV, while \citet{Walton10} used a self-consistent, relativistic 
reflection scenario with a full-covering absorber and successfully interpreted the data.
\citet{Gofford14} \& \citet{Matzeu16} suggested that partial-covering Compton-thick winds are
required to interpret the spectral variability shown in PDS~456, though latest 
work by \citet{Matzeu17} ruled out partial-covering absorption to be the physical 
mechanism behind short timescale variability, as the timescales are too short to be 
induced by variable absorption. Since the 2011 and 
2013 observations have not been tested using the relativistic reflection model, 
we include spectral analysis in this work though we mainly focus on timing analysis 
here.

\citet{DeMarco13} reported a negative lag of $\sim$400 s at $\sim3\times10^{-5}$ Hz, 
which was not significantly 
detected ($<1\sigma$), between the soft ($0.3-1$ keV) and hard ($1-5$ keV) 
energy bands using short \emph{XMM-Newton} observations. Based on the 
frequency-mass relation of \citet{DeMarco13}, the lag of PDS~456 should appear
at $\sim10^{-5}$ Hz and a frequency band including frequencies lower than $10^{-5}$ Hz 
needs to be probed to confirm detection. In this work, we use long 
\emph{Suzaku} observations to probe low Fourier frequencies which \emph{XMM-Newton} 
data cannot reach. We select the soft and hard bands, which are slightly different 
from the choice of \citet{DeMarco13}, based on results of spectral analysis.

In all the spectral analyses, which were performed using the XSPEC 12.8.2 package
\citep{Arnaud96}, the ``wilm" abundances \citep{Wilms00} and ``vern" 
cross section \citep{Verner96} were used. 
All of the errors quoted in the spectral analysis represent the 90 per cent confidence level,
while errors in the timing analysis represent 1$\sigma$ uncertainties.

\subsection{Spectral Analysis} \label{subsec_spectral}

We fit the FI XIS spectra in the $0.5-10.0$ keV energy range, excluding energies from $1.7-2.1$ keV due to
known calibration issues.  We fit the PIN spectrum over the $15.0-45.0$ keV energy band but only the 
spectrum of the 2007 observation was included. Note that there is mild spectral
variability in the 2013 observation. However, in this work we focus on the overall 
interpretation of the spectrum rather than the variations of warm absorbers, and we use
the combined 2013 spectrum for spectral fitting.
The data were
fit using the same model used in \citet{Walton13BareAGN}, which consists of a powerlaw and a 
relativistic disc reflection components, and an {\sevensize XSTAR} grid (the same one
used in \citealt{Walton10,Walton13BareAGN}) to model the 
absorption feature shown around the Fe K band caused by the ultrafast flow. Galactic 
absorption was accounted for using the {\sevensize TBABS} model in XSPEC. We use
the REFLIONX grid \citep{Ross05} to model reflected emission, and the RELCONV 
\citep{Dauser10} kernel to act on the reflected emission to account for relativistic effects. 
We assume the inner
edge of the accretion disc to extend down to the innermost stable circular orbit (ISCO),
and set the outer radius to be 400 $R_{\rm g}$ ($R_{\rm g}=GM/c^2$). The iron 
abundance of the accretion disc and the outflow is set to be identical.

In the spectral fitting we set parameters which are not expected to change significantly over
time to be the same in all data. These are, the column density of Galactic absorption 
$N_{\rm H}$, the iron abundance $A_{\rm Fe}$, the spin parameter $a^*$, and the 
inclination angle $i$.
\citet{Reeves09} have shown that the ultrafast outflow is persistent throughout the 2007 
and 2011 \emph{Suzaku} observations, with an outflow velocity of 0.25-0.3$c$. 
\citet{Matzeu16} examined the 2013 \emph{Suzaku} observation and found an outflow
velocity of $0.25^{+0.01}_{-0.05}c$. \citet{Nardini15} also reported a similar value of 
$0.25\pm0.01c$ using the long 2013-14 \emph{XMM-Newton} data, implying the 
ultrafast outflow in PDS~456 remains at constant velocity across several years. Hence
we link the redshift of the {\sevensize XSTAR} grid across all data as well. In order to
expedite the fitting process, we fit all data together to obtain the best-fitting `unchanged 
parameters' mentioned above, and then fit each observation individually with restricting
these parameters to vary within the best-fitting error range. We show the best-fitting
values of the `unchanged parameters' in Table \ref{spec}.

\begin{figure}
\begin{center}
\leavevmode \epsfxsize=8.5cm \epsfbox{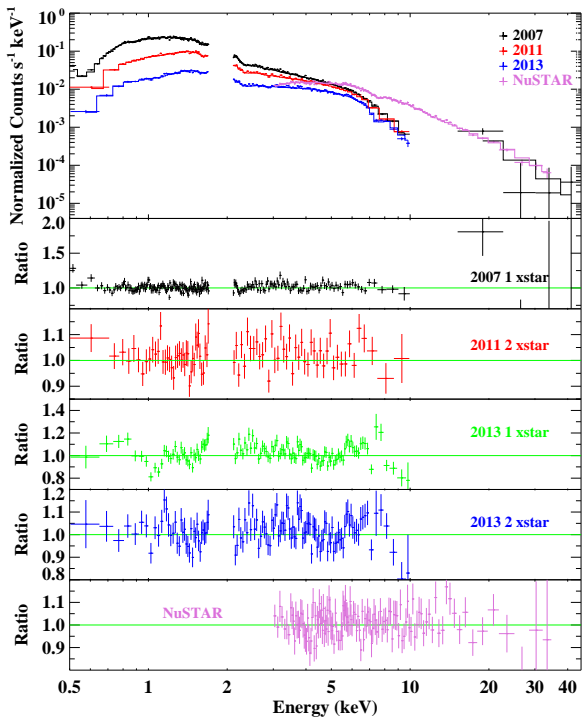}
\end{center}
\caption{Spectral fits to the \emph{Suzaku} and 2013 \emph{NuSTAR} data of PDS~456 in observed frame. The top panel shows 
the data used in this work and the lower four panels present the data/model ratios.
The 2007 spectrum only requires one {\sevensize XSTAR} component to account
for the ultrafast outflow, while the 2011 and 2013 data are better interpreted with
an additional low-ionized {\sevensize XSTAR} component. We 
show the fitting result of the 2013 observation, and improvement with the second
{\sevensize XSTAR} component in different panels. The last panel shows the fitting
result of the 2013 \emph{NuSTAR} observation, and it can be seen that the model fits the data very well.
}
\label{spec_fitting}
\end{figure}

\begin{figure}
\begin{center}
\leavevmode \epsfxsize=8.5cm \epsfbox{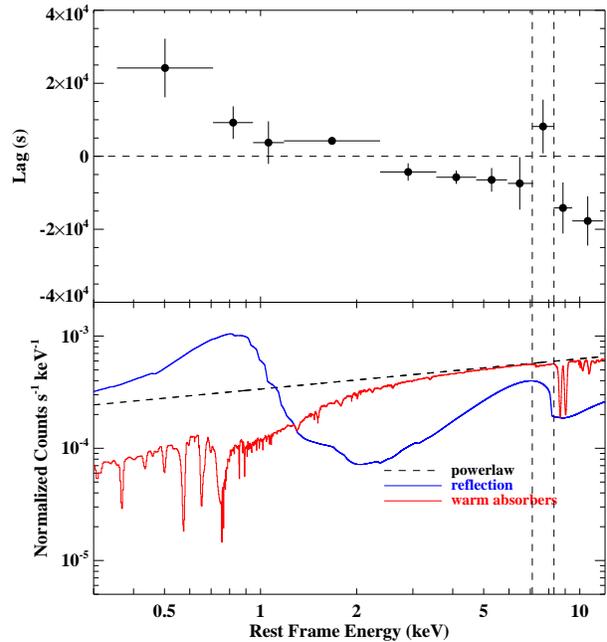}
\end{center}
\caption{{\it Top: }The lag-energy spectrum from all \emph{Suzaku} observations in the $[7.2-12.7]\times10^{-6}$ Hz frequency range. {\it Bottom:} The decomposed best-fitting spectral model of the 2013 \emph{Suzaku} observation in rest frame.  The vertical dashed lines are to guide the eye in associating the enhanced lag in the 7.1 -- 8.3 keV band with the blue-wing of the Fe K emission line in the best-fitting model.}
\label{model}
\end{figure}

The model results in a good $\chi^2/d.o.f.$ of 1131/1107 for the the 2007 observation
and the best-fitting parameters are shown in Table \ref{spec}. We obtain $\chi^2/d.o.f.$
= 831/788 and 1300/1124 for the 2011 and 2013 data, respectively. It seems there is a
mild curvature around 1 keV in the 2013 spectrum (see the second last lower panel in 
Fig. \ref{spec_fitting}) which causes residuals 
and the model could not give as good a fit as for the 2007 and 2011 data. The feature could be
caused by emission or absorption. We find that the curvature cannot be modeled by
a single emission/absorption line nor by an absorption edge. If including an ionized 
absorber (ABSORI in XSPEC) in the model, the fitting improves to $\chi^2/d.o.f.$
= 1171/1124, implying that the distortion in the 2013 spectrum is caused by a warm absorber.
Moreover, \citet{Nardini15} also identified a fully-covered warm absorber that affects the
spectrum below $<$ 2 keV in the 2013 \emph{XMM-Newton} observation.
We generate another {\sevensize XSTAR} grid (assuming $\Gamma=2$) with a lower 
ionization range (log $\xi=0.5-2$), to account for the effects of the warm absorber. 
Note that we set the covering fraction for both the high-ionization 
(to account for the ultrafast outflow) and low-ionization (to model the $<$ 2 keV spectrum 
affected by warm absorber) {\sevensize XSTAR} grids 
to be 100 per cent (fully-covered). The total model is now: TBABS*XSTAR$_{1}$*
XSTAR$_{2}$*(POWERLAW + RELCONV*REFLIONX). Since there is no resolved 
absorption lines to constrain the outflow velocity, we fix the redshift of the second 
{\sevensize XSTAR} grid at $z=0.184$, the redshift of PDS~456. Note that 
\citet{Reeves16} reported broad 
absorption lines at soft X-ray energies and found an outflow velocity of $\sim0.1-0.2c$. 
We do not probe the details of the second outflow here as the \emph{Suzaku} XIS 
data are not good enough to constrain its properties. We find that although the 
curvature does not seem to appear in the 2011 spectrum, the fitting improves a bit 
with the low-ionization {\sevensize XSTAR} grid. Best-fitting parameters of the model 
including two {\sevensize XSTAR} grids are list in Table \ref{spec} as well. We show
the fitting results and decomposed model in Figs. \ref{spec_fitting} \& \ref{model}. It can be seen that $0.3-0.8$
keV is dominated by reflection and $1-3$ keV is primarily contributed by the 
powerlaw component. We choose these energy bands as ``soft" and ``hard" bands
to perform the timing analysis.

It can be seen that no partial-covering absorbers are included in our model, yet
spectral variability can be explained by changes in the continuum and evolution of
the warm absorbers. The model fits the 0.5-10 keV energy band well. Wiggles around
$\sim$9 keV in observed frame are likely due to the K$\beta$ and K edge absorption 
from Fe XXVI \citep{Nardini15}. The effective areas of the \emph{Suzaku} XIS detector drop significantly
around this band, making detailed modeling difficult. The column density of the 
high-ionization warm absorber obtained in this work is much lower than that reported in 
\citet{Nardini15} ($N_{\rm H}=6.9^{+0.6}_{-1.2}\times10^{23}$ cm$^{-2}$). This might be due to different
assumptions made when creating the {\sevensize XSTAR} grid, and the high iron 
abundance obtained in this work. In addition, it is known that the UFO evolves slightly
over time, and the data they used were not simultaneous with those used in the present work.
The best-fitting column density of the less-ionized absorber ({\sevensize XSTAR$_{2}$}) is close to
that obtained in \citet{Nardini15}, but with a slightly higher ionization parameter 
($N_{\rm H}\sim (1.8-4.3)\times10^{21}$ cm$^{-2}$ and log $\xi=0.31\pm0.02$ in their work).
We also test the model on the 2013 \emph{NuSAR} data which was reduced following the same 
procedures in \citet{Nardini15}, and it fits the 3-35 keV energy band in observed frame very well 
(see the last panel of Fig. \ref{spec}). We obtain a good $\chi^2/d.o.f.$ of 348/342, and major 
parameters (photon index, ionization parameters and column densities of warm absorbers, etc.) 
are broadly consistent with those reported in \citet{Nardini15}. This indicates that our model can 
interpret the broadband NuSTAR data successfully.

We also replace the REFLIONX grid with the XILLVER \citep{Dauser10,Garcia14}
model to test if the results are consistent. We find that major line parameters such
as the spin parameter and iron abundance are broadly consistent. The XILLVER
model tends to give a higher photon index, a higher emissivity index and a lower
ionization parameter, which is likely caused by different assumptions made between
these two models. In general, REFLIONX and XILLVER give consistent results to
the data.
 
\subsection{Timing Analysis}

The Fourier phase lag between the hard and soft energy bands can be comptuted 
following the technique described in detail in \citet{Uttley14}. In traditional Fourier analysis, 
continuous light curves are required. However, the technique is not possible with 
\emph{Suzaku} light curves as the telescope was in a low Earth orbit with an orbital
period of $\sim$5760 s, causing frequent gaps in the data. Although \emph{Suzaku} 
data cannot be analysed by the traditional Fourier technique, the duration of an 
observation is around twice of its exposure time, making these data ideal to probe the 
low Fourier frequency domain which is desired when studying massive AGN. \citet{Zoghbi13} 
developed a method to determine the powerspectrum density and lags from 
non-continuous data, containing gaps.  This
method uses a maximum likelihood approach to calculate time lags by fitting 
the light curves directly. It has been shown by Monte Carlo simulations to give the 
same results as the traditional Fourier method. Details of how uncertainties are estimated
can be found in \citet{Zoghbi13}.

\begin{figure}
\begin{center}
\leavevmode \epsfxsize=8.5cm \epsfbox{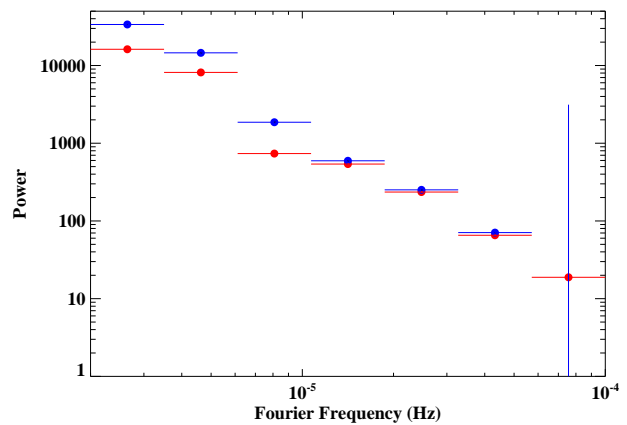}
\end{center}
\caption{The PSD of the soft ($0.3-0.8$ keV; 
$0.36-0.95$ keV in the rest frame; red data points) and hard ($1-3$ keV; $1.18-3.55$ keV 
in the rest frame; blue data points) light curves.
}
\label{psd}
\end{figure}

\begin{figure}
\begin{center}
\leavevmode \epsfxsize=8.5cm \epsfbox{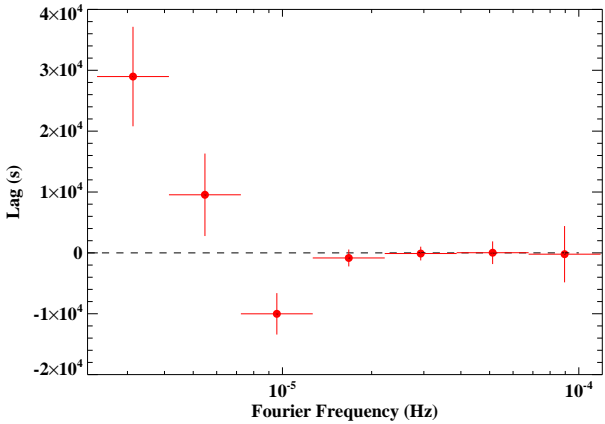}
\end{center}
\caption{The lag-frequency spectrum computed using $\sim$ 770 ks
of \emph{Suzaku} data. The lag is calculated between the soft ($0.3-0.8$ keV; 
$0.36-0.95$ keV in the rest frame) and the hard ($1-3$ keV; $1.18-3.55$ keV 
in the rest frame) energy bands, and a negative lag means that the soft band lags behind
the hard band. It can be seen that the most negative lag is $10000\pm3400$ s at 
$9.58\times10^{-6}$ Hz.
}
\label{lag_spec}
\end{figure}

The \emph{XMM-Newton} observations used by \citet{DeMarco13} only have 
exposures up to $\sim$ 90 ks. The longest observation used in 
this work was of $\sim$438 ks duration, and we generated the light curves 
using an 1 ks time bin, leading to a minimum Fourier frequency of 
$\sim2.3\times10^{-6}$ Hz and a maximum of $5\times10^{-4}$ Hz (the 
Nyquist frequency). Roughly even logarithmically-spaced 
frequency bins were used and we ensure at least 3 data points to fall within each
frequency bin. In Fig. \ref{psd} we show the power spectrum density (PSD) of the soft and hard
light curves. Fig. \ref{lag_spec} shows the lag-frequency spectrum between the 
soft ($0.3-0.8$ keV, dominated by disc reflection, $0.36-0.95$ keV in the rest frame) 
and the hard ($1-3$ keV, dominated by primary powerlaw emission, $1.18-3.55$ keV
in the rest frame) energy bands. The trend 
that lags switch from negative to positive (the hard flux lags behind the
soft) at low frequency has been seen in a number of previous timing studies 
\citep{Zoghbi11,Emm11,Kara13IRAS,Kara13,DeMarco13}, and again occurs in the lag-frequency
spectrum of PDS~456 though there is only one frequency bin due to the maximum duration of the lightcurves. 
We found the maximum negative lag where the soft band 
lags the hard by $10000\pm3400$ s to appear at $[7.2-12.7]\times10^{-6}$ Hz
(the amplitude of the soft lag and frequency have been corrected for cosmological redshift). 
\citet{Epitropakis16} indicated that too few Fourier frequencies within a bin
can  bias  the result, and there are six Fourier frequencies in this bin.
We also examine the long 2013-14 \emph{XMM-Newton} data. Nonetheless, the 
longest light curve in that observation is $\sim$ 138 ks and there is not enough
signal-to-noise ratio to probe the frequency band where the maximum negative lag
occurs. We only display results using the \emph{Suzaku} data.

\begin{figure}
\begin{center}
\leavevmode \epsfxsize=8.5cm \epsfbox{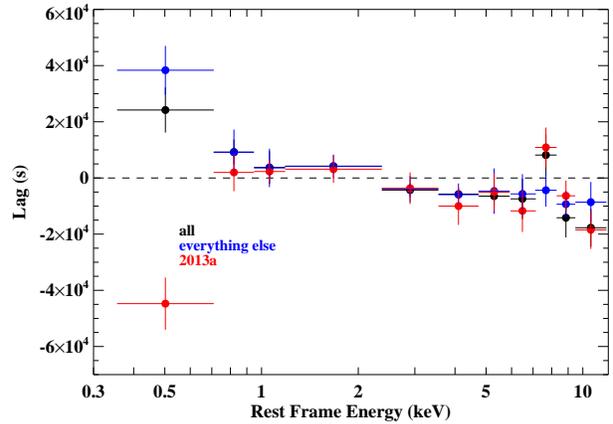}
\end{center}
\caption{The lag-energy spectrum of \emph{Suzaku} data.
The x-axis shows energy in rest frame. The spectrum shows the energy dependence
of the lags in the frequency range $[7.2-12.7]\times10^{-6}$ Hz, where the
maximum negative soft lag occurs based on the lag-frequency spectrum 
(Fig. \ref{lag_spec}). The black points represent results obtained using all data; the red
points were calculated using 2013a observation only; the blue points show results obtained
using all data except the 2013a observation.
}
\label{lag_energy}
\end{figure}

We then take a closer look at the result by generating a lag-energy spectrum to
examine how the lag evolves with energy. The lag is calculated between the light
curve in each energy bin and the light curve of the reference band. The reference 
band is the whole energy range ($0.3-10$ keV) excluding the energy bin of interest. 
For instance, the reference light curve for the $0.3-0.4$ keV band is 
produced by subtracting the $0.3-0.4$ keV light curve from the $0.3-10$ keV light 
curve. The reason to exclude the current band is to ensure the noise remains
uncorrelated between the bands (see \citealt{Zoghbi11}). Although the choice of reference band would cause the reference
light curve to be slightly different for each energy bin, Monte Carlo simulations 
showed there are only very mild differences ($\sim$ 1 per cent of systematic error).
Figures \ref{model} and \ref{lag_energy} show the lag-energy spectrum in the $[7.2-12.7]\times10^{-6}$ 
Hz frequency band. The lag begins to have a general negative trend starting at
around 3 keV in rest frame except the $7.1-8.3$ keV energy bin. The shape of the 
lag-energy spectrum is similar to that of IRAS~13224-3809 at high-flux states 
\citep{Kara13IRAS}.  From looking at the lag-energy spectra from individual observations, 
we find that the $7.1-8.3$ keV bin is dominated by the 2013a observation (see Fig. \ref{lag_energy}). 
This energy of this bin lines up with the peak 
of the Fe K line (see Fig.~\ref{lag_energy}), and may therefore be associated with an Fe K lag, though it cannot be confirmed 
based on current data. We note also that the lowest energy bin at $\sim0.5$ keV in the 2013a observation
shows a significant difference compared to the average lag-energy spectrum.  Changes in the shape of the reflection spectrum (such as the ionization parameter), can cause changes in the lag-energy spectrum \citep{chainakun15,chainakun16}.  Moreover, this observation also requires the strongest warm absorber which could affect the lag there (see Discussion).

\begin{figure}
\begin{center}
\leavevmode \epsfxsize=8.5cm \epsfbox{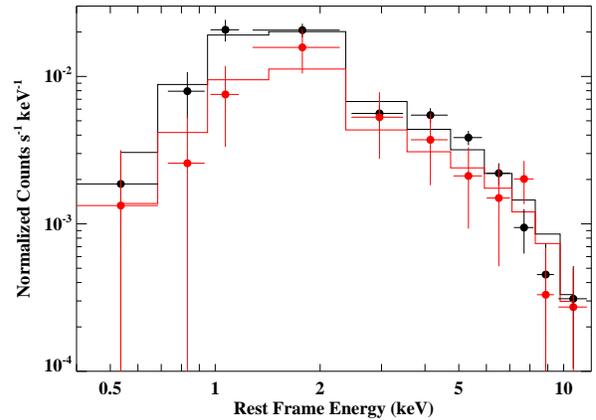}
\end{center}
\caption{The covariance spectrum for the frequency band $[7.2-12.7]\times10^{-6}$ Hz. 
Black points show the overall covariance spectrum, and
red ones the covariance spectrum of the 2013a (ObsID: 707035010) observation.
}
\label{cov}
\end{figure}

To investigate the variable spectrum further, we create the covariance spectrum, which 
is a measure of the absolute amplitude of correlated variations in count rate as a function of 
energy \citep{Wilkinson09,Uttley14}, following the method described in \citep{Uttley14}.
The covariance spectrum only picks out the correlated variations and allows direct 
comparison with the time-averaged spectrum. Fig. \ref{cov} shows the covariance spectrum of the 
$[7.2-12.7]\times10^{-6}$ Hz frequency band, where the soft lag is detected. Notably, 
the covariance spectrum of the 2013a observation shows an enhancement at the same 
energy as the tentative Fe K lag.  Overall, the covariance spectrum roughly follows a 
power-law with $\Gamma = 1.73$, though there are clear deviations away from it.

\section{Discussion} \label{discussion}

The spectrum of PDS~456 can be simply explained by relativistic reflection
with two full-covering absorbers, where one is associated with the ultrafast 
outflow, and the other related to another warm absorber with a lower 
ionization state and a lower outflow speed which cannot be tested 
by the \emph{Suzaku} data. We do not find any partial-covering absorbers that are 
required to interpret the data. The change in the continuum,
the column density and ionization state of the warm absorbers are sufficient
to explain the spectral variability between observations.

We find a soft lag of $10000\pm3400$ s at $9.58\times10^{-6}$ Hz via the
maximum likelihood method, which is the largest soft lag and the lowest
frequency to date in X-ray reverberation studies. Previous works do not probe
Fourier frequencies below $10^{-5}$ Hz because there are no continuous 
light curves that are long enough to reach the low frequency domain. We 
will compare the results with previous literature in the following.

\subsection{Mass-scaling Relation}

\begin{figure}
\begin{center}
\leavevmode \epsfxsize=8.5cm \epsfbox{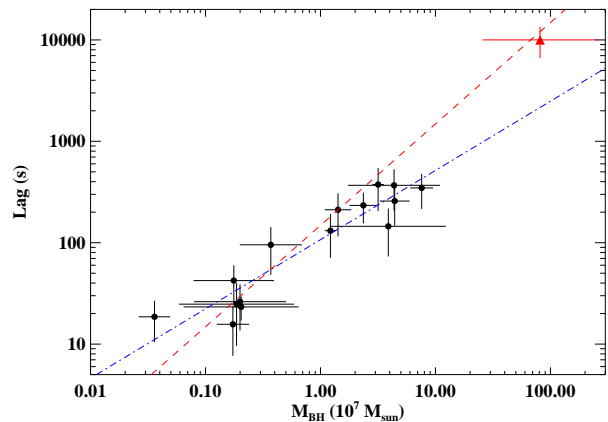}
\end{center}
\caption{Lag vs. black hole mass scaling relation from \citealt{DeMarco13} (black). 
The y-axis shows the most significant soft lags detected in AGN.
The red triangle on top represents the data point of PDS~456 measured in this work.
The blue dot dash line is drawn using the best-fitting relation of data from 
\citet{DeMarco13}. The red dash line presents the the light-crossing time at 
3 $R_{\rm g}$ as a function of mass.
}
\label{scaling_relation}
\end{figure}

We first compare our result with those obtained in \citet{DeMarco13}. In 
Fig. \ref{scaling_relation} we present the result of PDS~456 using a red 
triangle data point. It can be seen that PDS~456 lies close to the extrapolated
relation obtained by \citet{DeMarco13}. The data point of PDS~456 is 
slightly above the mass-scaling relation, which could be explained by the tentative correlation 
between lag and Eddington ratio \citep{Kara16}. PDS~456 is claimed to be accreting at 
Eddington or above and it is not surprising that it is slightly away from the
relation.  
In fact, the light travel distance converting from the 
amplitude of the soft lag is within 3 $R_{\rm g}$ (see the red dash line in Fig. \ref{scaling_relation}), 
which is as small as values
found in other AGN ($1-10$ $R_{\rm g}$). Previous work of \citet{DeMarco13} only probe Fourier
frequency in the range of $10^{-5}$ to $10^{-3}$ Hz and AGN with lower masses 
due to the use of continuous \emph{XMM-Newton} observations. The data point of
PDS~456 may deviate from the extrapolated best-fitting relation because of potential biases in
the \citet{DeMarco13} sample. The maximum length of the \emph{XMM-Newton} light curves
biases against finding large soft lags in massive black holes, potentially flattening the
relation.
Our result shows that the massive quasar PDS~456 follows the mass-scaling relation as well. 
Lags of sources in the De Marco sample and PDS~456 all imply that these signals 
are from areas very close to central black holes.
Although it is not known if all massive AGN follow the trend, the result of 
PDS~456 further supports the hypothesis that soft lags originate from the innermost 
areas of AGN.


\subsection{Modelling the Fe K$\alpha$ lag}
To better understand the expected Fe K$\alpha$ reverberation lag in the frequency range $[7.2 - 12.7]\times10^{-6}$~Hz, 
we use the models described in \citet{Cackett14} to calculate the transfer function for a point source at some height, $h$, 
above the black hole.  The best-fitting spectral parameters indicate an inclination of $65^\circ$ and a near maximally 
spinning black hole.  We therefore calculate the transfer function assuming $i = 65^\circ$ and $a = 0.998$.  We also 
assume a $h = 10$~GM/c$^2$, close to what was found for NGC 4151 in \citet{Cackett14}.  The spectral fits also indicate 
a reflection fraction (reflection flux / power flux) of approximately 0.6 at the peak of the iron line in the 2013 observations 
(it is lower in the other observations).  Furthermore, we assume $\log M = 8.91$ \citep{Zhou05}.  

\begin{figure*}
\includegraphics[width=0.9\textwidth]{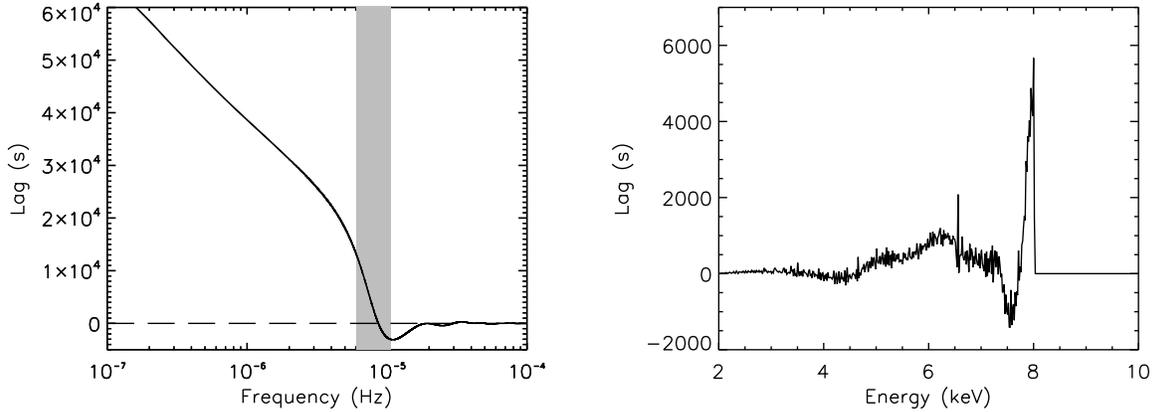}
\caption{{\it Left:} The lag-frequency spectrum for the energy-averaged iron line, assuming $i = 65^\circ$, $h = 10$~$GM/c^2$, $a=0.998$ and $\log M = 8.91$.  The grey shaded region shows the frequency range where the soft lag is detected in PDS 456, and over which the lag-energy spectrum (right-panel) is calculated. {\it Right:} The lag-energy spectrum over the frequency-range $[7.2 - 12.7]\times10^{-6}$~Hz, assuming a reflection fraction of 0.6, and parameters listed previously.}
\label{fig:fekmodel}
\end{figure*}

The aim here is not to fit the data, but just to explore what might be expected to be observed.  With the parameters given 
above we produce the lag-energy spectrum at the frequency of the soft lag in Figure~\ref{fig:fekmodel}.  The lag-frequency 
spectrum for the energy-averaged iron line is also shown.  From the lag-frequency spectrum we see that for such a massive 
black hole the frequency-range $[7.2 - 12.7]\times10^{-6}$~Hz is right where phase-wrapping is occurring.  This means that 
we expect to see only reverberation from the regions of the disc with the shortest path length differences.  With a high 
inclination angle of $65^\circ$ this will be close to the black hole on the near-side of the accretion disc.  We therefore expect 
to see lags from only the most red - and blue-shifted parts of the line (see Cackett et al. 2014 for detailed descriptions).  
What is striking about the resultant lag-energy spectrum in Figure~\ref{fig:fekmodel} is the narrow peak in the lags in the 
7$-$8 keV range, coming from the most blue-shifted emission.  The general shape of the lag-energy spectrum is very similar 
to what is observed in PDS~456 -- a peak in only the $\sim$7$-$8 keV range and much smaller lags, with little energy-dependence, elsewhere.

We have not calculated a grid of models covering a range of parameters in order to fit the data properly. We do note, 
however, that given how the frequency range is right where the phase-wrapping crosses zero, small changes in the 
assumed mass, height, and inclination can have a big effect on the lag-energy spectrum.  If the height of the corona 
changes with flux (as suggested by spectral fitting, e.g., \citealt{Fabian12}) then this could explain why only the 2013a 
observation shows a significant lag in the 7.1$-$8.3 keV range.

\subsection{Influence of the Outflow}

\citet{Emm11} carried out timing analysis on MCG-6-30-15 using the $0.5-1.5$ keV
and $2-4$ keV light curves. The $0.5-1.5$ keV energy band of MCG-6-30-15 is 
seriously modified by warm absorbers \citep{Chiang11} and the soft lag obtained 
from the data might be caused by scattering/absorbing material. Nevertheless,
\citet{Emm11} found that the soft lag of MCG-6-30-15 is most likely caused by 
reflection by testing various models. Reverberation lags can still be detected even
when the timing analysis involves energy bands which are strongly affected by
warm absorbers, but strong absorption does inhibit the measurement of time lags.
\citet{Silva16} investigated data of NGC~4051 and found that warm absorbers do 
produce time lags on the order of $\sim$ 50 s or more at lower frequencies, which 
is of a longer timescale than the reverberation lags in NGC~4051 and hence does not pollute
the measurement. The time delay is related to the response of the gas to changes 
in the ionizing continuum and is dependent on the density of the warm absorber. 
However, only warm absorbers within a certain distance ($10^{15}$ cm to 
$10^{16.5}$ cm in the case of NGC~4051, $\sim$3900-123000 $R_{\rm g}$) 
matter. Those close to the ionizing source 
respond immediately, and gas further out is unable to respond to create a visible time 
delay. In NGC~4051 the component with a mild ionization parameter of log $\xi\sim$ 
2.9 and $N_{\rm H}=(1.2\pm0.2)\times 10^{21}$ cm$^{-2}$ contributed the majority 
of the soft lag, while the highly-ionized (log $\xi\sim$ 3.7) and the least ionized 
(log $\xi\sim$ 0.4) components cause little differences. The warm absorbers
found in PDS~456 are the same type which contributes least in NGC~4051, but 
column densities are higher than those found in NGC~4051. Assuming that
the position of warm absorbers scale with black hole masses, those in PDS~456
should not produce negative lags which coincide with reverberation lags. Nonetheless,
detailed analysis is still required to completely rule out the possibility that the 
time lag detected in PDS~456 is partly produced by variable warm absorbers, as
these outflows are complex and can be case dependent.


\citet{Kara16} examined a sample of Seyfert galaxies, including bare and 
partially obscured sources, and found $\sim$ 50 per cent of them exhibit iron K reverberation. 
An iron K lag was not detected in some sources with complex absorption such as 
MCG-6-30-15, but seen in several sources with warm absorbers. IRAS13224-3809,
which is known to have ultrafast outflow appearing at the iron K band \citep{Parker17}, does reveal 
an iron K lag. The presence of warm absorbers does not seem to destroy iron K lags.
Warm absorbers do not seem to leave signatures in the lag-energy spectrum 
(Fig. \ref{lag_energy}) of PDS~456. The negative trend is disrupted by the $7.1-8.3$ 
keV energy bin in rest frame, where neither of the warm absorbers would affect 
significantly (the ultrafast outflow locates at $8-9$ keV rest frame, and the second 
warm absorber casts effects on low-energy band as seen in Fig. \ref{model}).
Nonetheless, the $7.1-8.3$ keV positive lag cannot be confirmed as an iron K lag
either. Without enough evidence, the X-ray lag shown in PDS~456 cannot be 
concluded to be reverberation lag.

\citet{Kara15} showed that the iron K reverberation lag cannot be 
found in NGC~1365 in orbits that are highly obscured. They also found a low-frequency 
soft lag (instead of hard lag as found in other AGN) in a less-absorbed orbit, which can 
be explained as a transient phenomenon. There is evidence of a neutral eclipsing 
cloud in NGC~1365, and the low-frequency soft lag is caused by a change in 
column density, which can be interpreted as the eclipsing cloud moving out the line 
of sight. In the case of PDS~456, \citet{Matzeu16} reported significant short-term 
X-ray spectral variability on timescales of $\sim$100 ks in the 2013 observation, 
and suggested that the spectral variability can be accounted for by variable covering 
of the clumpy absorbing gas along the line of sight. If this is true, we should be 
able to see signatures in the lag-frequency spectrum of the 2013 observation.
The lag-frequency spectrum from the combined 2013 observation shows no
significant difference from Fig. \ref{lag_spec}.  However, we do see that when looking at only
the 2013a lag-energy spectrum in the $[7.2-12.7]\times10^{-6}$ Hz frequency range, that there
is a change in the lag in the lowest energy bin (which would be effected most by neutral absorption).
But, the lag changes in the opposite sense from what would be expected by the presence 
of an eclipsing cloud with high column density in PDS~456 during that observation only. 
This does not 
reject the possibility of partial-covering Compton-thick absorbers, which could be
located further from the centre and do not cast effects on lag measurement. 
However, recent X-ray micro-lensing studies have shown that the X-ray emitting region of 
quasars is compact (less than $\sim$10 gravitational radii $R_{\rm g}$, 
\citealt{Dai10,Chartas12}). If clumpy partial-covering absorbers are not in the 
innermost regions of AGN, they are more likely to fully cover the source. 
We already show in Section \ref{subsec_spectral} that a simple 
relativistic reflection model with Compton-thin, fully-covered warm absorbers 
can interpret the spectrum of PDS~456 without difficulty. In the timing analysis
we find no evidence to contradict the model.

Based on present work, we did not find evidence for a high-column eclipsing cloud such as 
that found in NGC~1365 to be present in PDS~456. Although warm absorbers can affect the 
 lag measurement, those with low column density do not seem to contribute 
significantly to timing analysis. Combining both spectral and timing analyses, the 
time lag in PDS~456 is more likely to be an X-ray reverberation lag. Nevertheless, given
that there is only tenative 
evidence of an iron K lag, the possibility that the time 
lag was generated by warm absorbers cannot be completely ruled out.

\section{Conclusions}

We present both the spectral and timing analysis of PDS~456 using $\sim$
770 ks of archival \emph{Suzaku} data. We find that the spectrum can be
simply explained by relativistic reflection with two full-covering warm 
absorbers. One of them is the persistent, highly-ionized ultrafast outflow; 
the other is less ionized and started to appear in the 2011 observation.
No partial-covering absorbers were required to interpret spectral variability, 
and we did not find effects caused by these absorbers in the lag-frequency 
spectrum either. Based on the spectral analysis, PDS~456 harbours
a rapidly-spinning black hole of $a^{*}>0.99$. Furthermore, we find a soft 
lag of $10000\pm3400$ s at $9.58\times10^{-6}$ Hz, which matches the 
scaling relation and indicates an origin within 3 $R_{\rm g}$. More data
are required to confirm the relation for massive AGN with $M>10^{8} M_{\odot}$,
but our result further supports the scenario that soft lags originate from 
the innermost regions of AGN. The results show that the relativistic
reflection model is self-consistent and reveal the robustness of probing
areas near the central engine using this model.

\section*{Acknowledgements}
This work was greatly expedited thanks to the help of Jeremy Sanders in
optimizing the various convolution models. We thank Phil Uttley for useful
discussions. EMC gratefully acknowledges support from the NSF through
CAREER award number AST-1351222. CSR thanks NASA for support 
under grant NNX15AU54G. ACF acknowledges ERC Advanced Grant 340442.

\bibliographystyle{mn2e_uw}
\bibliography{pds456}

\label{lastpage}
\end{document}